\documentclass[aps,prb,citeautoscript,floatfix,twocolumn,superscriptaddress,bibnotes]{revtex4-1}
\usepackage{physics}
\usepackage{graphicx}
\usepackage{xcolor}
\usepackage[pdftex,bookmarks=true,bookmarksopen,bookmarksnumbered,
                colorlinks,
                linkcolor=blue,
                citecolor=blue]{hyperref}
                
\usepackage{cleveref}
\usepackage{lipsum}
\usepackage{siunitx}

\usepackage{colortbl}
\usepackage{float}
\usepackage{newfloat}
\usepackage{pgfplots}
\usepackage{textgreek}
\pgfplotsset{compat=newest}
\usepgfplotslibrary{colorbrewer,patchplots}
\pgfplotsset{colormap/hot2}
\usepackage{standalone}
\graphicspath{
  {figures/}
}

\usepackage{pgfplotstable,booktabs}

\newcommand{\nv}{NV$^-\:$}

\begin{document} 

\title{Coherent control of \texorpdfstring{NV$^-$}{NV-} centers in diamond in a quantum teaching lab}

\author{Vikas K. Sewani}
\email[]{v.sewani@student.unsw.edu.au}
\affiliation{
Centre for Quantum Computation and Communication Technology, School of Electrical Engineering and Telecommunications, UNSW Sydney, Sydney, New South Wales 2052, Australia
}

\author{Hyma H. Vallabhapurapu}
\affiliation{
Centre for Quantum Computation and Communication Technology, School of Electrical Engineering and Telecommunications, UNSW Sydney, Sydney, New South Wales 2052, Australia
}

\author{Yang Yang}
\affiliation{
Centre for Quantum Computation and Communication Technology, School of Electrical Engineering and Telecommunications, UNSW Sydney, Sydney, New South Wales 2052, Australia
}

\author{Hannes R. Firgau}
\affiliation{
Centre for Quantum Computation and Communication Technology, School of Electrical Engineering and Telecommunications, UNSW Sydney, Sydney, New South Wales 2052, Australia
}

\author{Chris Adambukulam}
\affiliation{
School of Electrical Engineering and Telecommunications, UNSW Sydney, Sydney, New South Wales 2052, Australia
}

\author{Brett C. Johnson}
\affiliation{
Centre for Quantum Computation and Communication Technology, School of Physics, University of Melbourne, Melbourne, Victoria 3010, Australia
}

\author{Jarryd J. Pla}
\affiliation{
School of Electrical Engineering and Telecommunications, UNSW Sydney, Sydney, New South Wales 2052, Australia
}

\author{Arne Laucht}
\email[]{a.laucht@unsw.edu.au}
\affiliation{
Centre for Quantum Computation and Communication Technology, School of Electrical Engineering and Telecommunications, UNSW Sydney, Sydney, New South Wales 2052, Australia
}

\begin{abstract}
The room temperature compatibility of the negatively-charged nitrogen-vacancy (NV$^-$) in diamond makes it the ideal quantum system for a university teaching lab. Here, we describe a low-cost experimental setup for coherent control experiments on the electronic spin state of the NV$^-$ center. We implement spin-relaxation measurements, optically-detected magnetic resonance, Rabi oscillations, and dynamical decoupling sequences on an ensemble of \nv centers. The relatively short times required to perform each of these experiments ($<$10 minutes) demonstrate the feasibility of the setup in a teaching lab. Learning outcomes include basic understanding of quantum spin systems, magnetic resonance, the rotating frame, Bloch spheres, and pulse sequence development.

\end{abstract}

\maketitle

\section{Introduction}
The arrival of the second quantum revolution -- technologies that exploit coherent quantum phenomenon, such as quantum computation, quantum communication, and quantum sensors -- has elevated quantum physics from fundamental research to an applied science.\cite{Dowling2003} A thorough understanding of quantum mechanics is not only desirable but quite often demanded of university graduates entering the job market in this field. Theoretical skills are easily taught with pen and paper, by deriving analytical solutions, or with a few lines of computer code running numerical simulations, but real hands-on experience in manipulating a quantum system is, unfortunately, much more difficult to convey. This is partly due to the sensitivity of quantum systems to environmental disturbances that often demands operation at cryogenic temperatures or inside vacuum chambers, and partly due to the high cost of specialized equipment.

The NV$^-$ center in diamond is an especially advantageous quantum system for a teaching lab as its electron spins can be initialized, controlled, and read out at room temperature in ambient atmosphere,\cite{Doherty2013} strongly reducing the complexity of the experimental apparatus. Hence it has previously been recommended for teaching lab setups for magnetic resonance and magnetometry by Zhang \textit{et al.} in Ref.~\onlinecite{Zhang2018}, and is even available as a commercial system.\cite{Qutools} While the experimental apparatus required to extend the experiments to coherent control is more complex, it has been thoroughly described in literature by Bucher \textit{et al.} in Ref.~\onlinecite{Bucher2019}. However, coherent control has not been demonstrated in a convenient and cost-effective fashion targeted at a university teaching lab.

Here, we describe a teaching lab setup that allows students to learn about fundamental concepts of quantum mechanics by coherently controlling the electronic spin state of the NV$^-$ center in diamond. In fact, students can adapt and design their own control sequences and experiments. The presented setup is robust to its environmental conditions, such that it can even be operated in broad daylight on a normal desk, and does not require the carefully controlled environment of a research laboratory. Finally, the setup can be assembled for a total cost of less than USD 20k.

This paper is organized as follows: In the proceeding section (Section~\ref{sec_learn}) we will give an overview of the learning outcomes that can be conveyed with these experiments. In Section~\ref{sec_NV} we will give a short introduction to the NV$^-$ center in diamond, explaining how initialization, control, and readout of the electronic spin states is achieved. Section~\ref{sec_Eq} provides a detailed description of the experimental setup. More specifically, Subsection~\ref{sec_sample} describes the diamond sample we use, Subsection~\ref{sec_Op} describes the optical part of the setup, Subsection~\ref{sec_MW} describes the equipment required for delivery of a microwave (MW) field, and Subsection~\ref{sec_LI} describes the signal detection scheme using a lock-in amplifier in combination with digital pulse sequences. Finally, Section~\ref{sec_Ex} describes the various experiments that can be conducted with this setup, with special attention to the quantum mechanical concepts that these experiments convey.

\section{Learning outcomes}\label{sec_learn}
At UNSW Sydney, the experiments described below have been incorporated into a course targeted at the 4th-year undergraduate level. At the time of writing, we have thus far run the course for one semester in which students have been able to successfully perform all experiments, and demonstrate the learning outcomes. A total of 9 students were enrolled in that semester, and were divided into groups of 2-3. Each group had 2 hours per week to perform the experiments, for 4 weeks in total. Due to the number of groups, two identical setups were built and were operated simultaneously during the lab times.
During the course of the labs, it was essential for students to demonstrate links between experimental results, and both quantitative and qualitative understanding of quantum theory. More specifically, these labs encourage students to achieve the following: 
\begin{enumerate}
  \item Understand the \nv center structure, the spin initialization and readout procedure, and the spectral features.
  \item Incrementally develop pulse sequences with increasing complexity, i.e. starting from purely optical spin-dynamics for $T_{1}$ measurements, to complex dynamical decoupling sequences for $T_2$ measurements.
  \item Understand the rotating frame, magnetic resonance, two-axes control, and be able to follow the spin orientations along the Bloch sphere during pulse sequences.
  \item Understand the incremental steps required to implement dynamical decoupling pulse sequences, such as finding a spin transition in the spectrum, performing Rabi oscillations, and calibrating $\mathrm{\pi}$-pulse lengths. Due to the short experimental run times, students can often perform all experiments in a single 2-hour lab once they have obtained the expertise.
\end{enumerate}

Concepts demonstrated with these experiments are transferable to other quantum systems,\cite{Humble2019} like electron spin qubits confined to donors or quantum dots,\cite{Ladd2018,Zhang2018qubits} nuclear spin qubits,\cite{Vandersypen2005} superconducting qubits,\cite{Clarke2008} atoms in ion traps,\cite{Haffner2008} and magnetic resonance imaging (MRI).\cite{Boretti2019} Literature that is readily available on these systems frequently benchmarks quantum devices using the same methods as described below. Performing such experiments has, up until now, often only been available to research students in an expensive research lab.

\section{The \texorpdfstring{NV$^-$}{NV-} center in diamond}\label{sec_NV} 
\begin{figure} 
	\centering 
		\includegraphics[width=\columnwidth]{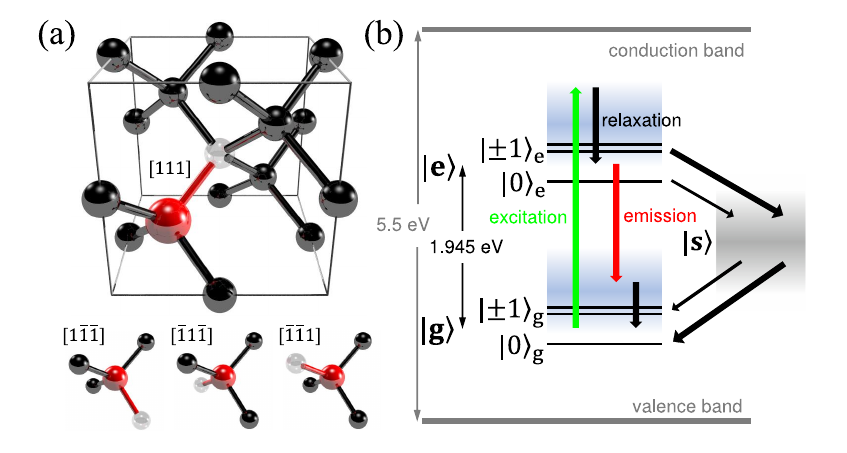}
		\caption{
		(a) Atomic structure of the NV center in diamond with a substitutional nitrogen atom (red) adjacent to a vacant lattice site (transparent). The depicted NV center is oriented along the [111] lattice direction, however due to the tetrahedral structure of the crystal lattice, the NV center axis can also be oriented along the [1$\bar{1}\bar{1}$], [$\bar{1}1\bar{1}$], and [$\bar{1}\bar{1}1$] lattice directions.
		(b) Energy levels and transitions of the NV$^-$. The blue shaded regions represent continua of orbital and vibrational states. Thick black arrows represent high probability transitions, while thinner arrows represent low probability transitions. The grey shaded region represents the intermediate singlet states. 
		}
		\label{fig_NV}
\end{figure}

The \nv center in diamond consists of a substitutional nitrogen atom adjacent to a vacant lattice site, as schematically shown in Figure~\ref{fig_NV}(a). It can exist along 4 different crystallographic orientations ([111], [1$\bar{1}\bar{1}$], [$\bar{1}1\bar{1}$], and [$\bar{1}\bar{1}1$]), which are in principle equivalent, but lead to different alignments of their \nv center axes with respect to an externally applied static or oscillating magnetic field. 

The static Hamiltonian of the NV$^-$ center ground state, neglecting any interactions with nuclear spins or spin-strain interactions, and assuming that the \nv-axis is oriented along the z-direction, is given by:
\begin{equation}
    \label{eqn:hamiltonian}
    {\cal H}_0=D{\rm S}_{\rm z}^2+\gamma_{e}B_{0}{\rm S}_{u},
\end{equation}
where $D=2.87$~GHz is the zero-field splitting of the ground state, $\gamma_{e}=28$~GHz/T is the electron gyromagnetic ratio, $B_{0}$ is the magnetic field applied along an arbitrary direction $\vec{u}$, and S$_{{\rm x,y,z,}u}$ are the spin matrices for $S=1$ along the x,y,z-axes and the $\vec{u}$-direction.\cite{Abe2018} In this paper, we will treat ${\cal H}_{0}$ in units of frequency, as this is more relevant for experiments. 

\begin{figure*} 
	\centering 
		\includegraphics[width=\textwidth]{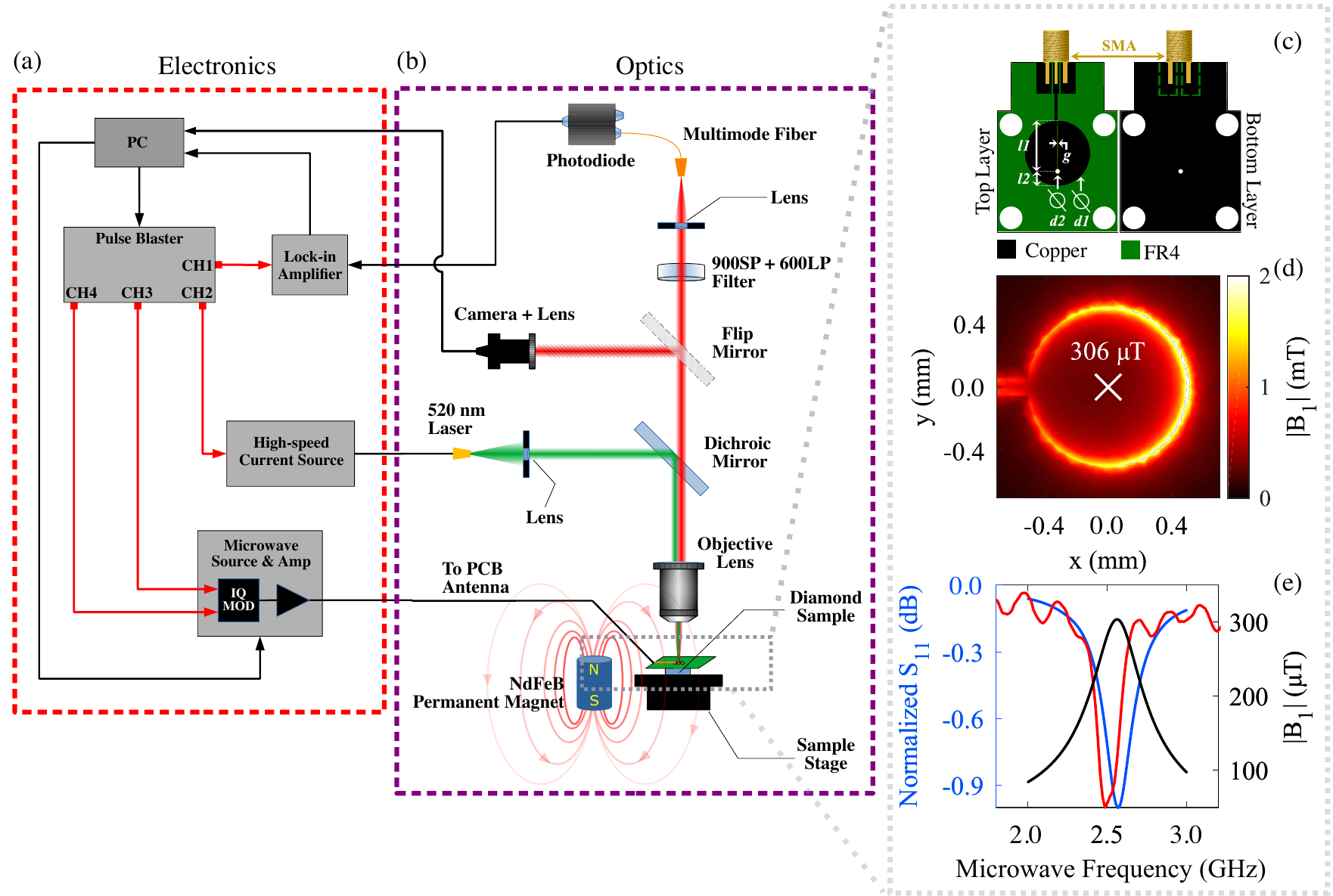}
		\caption{(a) Electronics setup. (b) Optics setup. (c) CAD drawing of the PCB antenna designed to deliver the oscillating magnetic field $B_{1}$ to the \nv spins. The geometric parameters are $d1=14$~mm, $d2=1.0$~mm (for laser pass-through), $l1=10.9$~mm, $l2=3.1$~mm, and $g=0.1$~mm. The bottom layer copper is shorted to the ground/shield of the SMA connector (not shown). The design is adapted from Ref.~\onlinecite{Sasaki2016}, and commercially fabricated on a 0.254~mm thick FR4 substrate with 36~\textmu m of copper thickness on each side. (d) Magnitude of the magnetic field $\abs{B_1}$ at +24~dBm of excitation at 2.57~GHz, around the 1~mm hole. Simulations were performed in CST Microwave Studio. (e) Comparison of simulated (blue) and measured (red) power reflected from the antenna (S$_{11}$), and simulated magnetic field magnitude $\abs{B_1}$ (black).}
		\label{system_arch}
\end{figure*}

In order to achieve magnetic spin resonance and coherent control, we need an oscillating magnetic field $B_{1}$ that induces transitions between the spin sub-levels. For example, a resonant $B_{1}$ field enables us to controllably rotate the ground state spin states from $\ket{0}$ to $\ket{\pm1}$ and back (termed Rabi oscillations). This field is usually created at the position of the \nv centers using a MW signal generator and an antenna (discussed further in Section \ref{sec_Eq}). We include the oscillating magnetic field in the Hamiltonian as a time-dependent term given by:
\begin{equation}
    \label{eqn:hamiltonian_mw}
    {\cal H}_{1}=\gamma_{e}B_{1}{\rm cos}(2\pi\nu_{1}t){\rm S}_{v},
\end{equation}
where $B_{1}$ is the magnitude of the oscillating magnetic field at frequency $\nu_{1}$ applied along an arbitrary direction $\vec{v}$, and S$_{v}$ is the spin matrix along the $\vec{v}$-direction. 

Due to the exact way the sample is mounted in the setup, there can be an angle between the static magnetic field $B_0$, the oscillating magnetic field $B_1$, and the direction in which the crystal field acts (defined as the z-direction in Equation \ref{eqn:hamiltonian}), as for example indicated in Fig.~\ref{odmr_rabi}(d). In fact, due to the 4 possible orientations of the NV$^-$ center axis in the tetrahedral crystal, NV$^-$s with different orientations will, intrinsically, have different Zeeman splittings and Rabi frequencies.

The experiments described in this paper will only be concerned with the spin physics of the ground state described above, denoted $\ket{\mathrm{g}}$ in Figure~\ref{fig_NV}(b). However, in order to describe the spin initialization and readout mechanisms, we must consider the optically excited state $\ket{\mathrm{e}}$ and the intermediate singlet states $\ket{\mathrm{s}}$, also depicted in Figure~\ref{fig_NV}(b). The ground state $|{\rm{g}}\rangle$ and excited state $|{\rm{e}}\rangle$ are both spin-carrying states with $S=1$. In $|{\rm{g}}\rangle$ at $B_{0}=0$ T, the $\ket{\pm1}_{\mathrm{g}}$ states are at $\nu_D=2.87$~GHz higher energy than the $\ket{0}_{\mathrm{g}}$ state due to the zero-field splitting $D$. Green laser light at $\lambda=520$~nm or $\lambda=532$~nm can excite the NV$^-$ electrons from the ground state $|{\rm{g}}\rangle$ into the continuum of orbital and vibrational excited states (upper blue shaded region) above the excited state $|{\rm{e}}\rangle$ (referred to as off-resonant excitation), from which they rapidly relax into $|{\rm{e}}\rangle$. From there the electrons can decay radiatively, either directly to $|{\rm{g}}\rangle$ by emitting a photon at $\lambda_{\rm{ZPL}}=637$~nm into the zero-phonon-line (ZPL), or by simultaneously emitting a phonon and a photon of longer wavelength into the phonon-sideband (PSB). In either case, there is a high probability that the spin state of the electron will remain unchanged during this optical cycling, due to spin-conservation. Alternatively, the NV$^-$ electrons can decay non-radiatively via the singlet state $|{\rm{s}}\rangle$. Decay from $|{\rm{e}}\rangle$ to $|{\rm{s}}\rangle$ is favoured by the $\ket{\pm1}_{\mathrm{e}}$ states compared to the $\ket{0}_{\mathrm{e}}$ state (as indicated by the thicker arrow), and decay from $|{\rm{s}}\rangle$ to $|{\rm{g}}\rangle$ favours the $\ket{0}_{\mathrm{g}}$ spin orientation of $|{\rm{g}}\rangle$. Overall, these transition rules have 2 effects: \\
1. Electrons cycling between the $\ket{0}_{\mathrm{g}}$ and $\ket{0}_{\mathrm{e}}$ states emit $\sim 30$\% more photons than those cycling between the $\ket{\pm1}_{\mathrm{g}}$ and $\ket{\pm1}_{\mathrm{e}}$ states. This provides a readout mechanism for the ensemble electronic spin state. \\
2. With continuous laser excitation, electrons cycling between $\ket{\pm1}_{\mathrm{g}}$ and $\ket{\pm1}_{\mathrm{e}}$ states will over time populate the $\ket{0}_{\mathrm{g}}$ state. The electrons can therefore be spin-initialized into the $\ket{0}_{\mathrm{g}}$ state.\cite{Doherty2013}

\section{Equipment}\label{sec_Eq}
In this section we provide details about the experimental setup. All optical and electrical components were purchased off-the-shelf, while the printed circuit boards for the antenna [see Fig.~\ref{system_arch}(c)] and the laser current driver (more information in Appendix~\hyperref[AppB]{B}) can be commercially manufactured (\textit{Eagle CAD} files are included in the supplementary material).\cite{PCBFiles} A detailed list of all parts with part numbers and recent prices can be found in Appendix~\hyperref[AppA]{A}.

\subsection{Diamond sample}\label{sec_sample}
The diamond sample is a single crystal Type 1b $\langle111\rangle$-oriented high-pressure, high-temperature (HPHT) diamond obtained from \textit{Sumitomo}. In order to obtain a high \nv concentration, the diamond was additionally electron-irradiated in-house with a density of $10^{18}$~electrons/cm$^2$, and annealed in vacuum at 900$^{\circ}$C for 2 hours. The electron irradiation significantly increases the PL intensity but reduces the coherence times,\cite{Rubinas_2018} which can be a limiting factor for some more advanced measurement protocols. All experiments measure the photoluminescence (PL) from a large ensemble of \nv centers with all four possible orientations. While the sample used in this manuscript requires some processing, a $\langle100\rangle$-oriented HPHT diamond sample with high \nv center density of a few ppm can be purchased from \textit{Scimed},~\cite{Scimed} and a $\langle100\rangle$-oriented chemically-vapour-deposited (CVD) diamond sample (available from \textit{Element Six})~\cite{Element6} is also sufficient for the experiments described herein and will contain a measurable number of \nv centers, without any additional processing. In Appendix~\hyperref[AppC]{C}, we show a comparison between the HPHT diamond that was used for the experiments in this manuscript and the commercially available CVD diamond from \textit{Element Six} for a zero-field spin resonance experiment (as in Section~\ref{sec_odmr}) and a Rabi experiment (as in Section~\ref{sec_Rab}). The CVD sample has a $\sim1000$ times lower PL signal, and while the spin signal decreases proportionally, we obtain high-quality data with only $\sim10$ times longer measurement times. In contrast, the CVD sample has a much longer coherence time than the HPHT sample that allows observation of effects originating from coupling to nuclear spins (not discussed in this work). The effect of the different crystal orientations with respect to the diamond surface on our experiment is described in Section~\ref{sec_Rab} and Endnote~\onlinecite{B1Dir}.

\subsection{Optical setup}\label{sec_Op}
The setup was designed with the goal of keeping the optical components to a minimum while ensuring ease of alignment. This way, the setup can be easily incorporated onto a small optical breadboard (Thorlabs MB3045/M) for portability, and fully covered with an enclosure (Thorlabs XE25C7/M) to constrain laser scattering.

A schematic of the optics part of our setup is shown in Figure~\ref{system_arch}(b). We use a single mode, fiber-coupled 520~nm green laser diode (Thorlabs LP520-SF15) as our excitation source. The beam is first collimated using an aspheric lens (Thorlabs C280TMD-A), and then reflected off a 550~nm long-pass dichroic mirror (Thorlabs DMLP550) and onto a microscope objective (Olympus MS Plan 50x/0.80NA). The objective lens focuses the laser to a focal spot of $\sim 1$~\textmu m on the diamond sample, which results in an excitation volume over which $B_{\mathrm{0}}$ and $B_{\mathrm{1}}$ field are both relatively constant. The diamond sample is mounted on a 3-axis translation stage (Thorlabs MBT616D/M). Between the objective lens and the diamond sample, we place a printed circuit board (PCB) MW antenna to create the strong oscillating magnetic field ($B_{1}$ field) for spin control. The antenna structure includes a 1~mm diameter hole, through which both the excitation laser, and the PL emission from the diamond sample can pass. The design of the antenna is discussed in more detail in Section~\ref{sec_MW}. 

The PL signal from the diamond sample is transmitted through the dichroic mirror and into the detection part of our setup. Here, a flip-mirror (Thorlabs PF10-03-P01 on Thorlabs TRF90/M) allows the signal to follow one of two paths: \\ 
1. Through an achromatic doublet (Thorlabs AC254-125-A) onto a CMOS camera (Thorlabs DCC1545M) that allows for alignment of the sample and focusing of the laser using the 3-axis stage. A 600~nm long pass filter (Thorlabs FEL0600) and a 900~nm short pass filter (Thorlabs FES0900) can be added here to view the PL signal instead of residual laser light. \\ 
2. Via two silver-coated mirrors (Thorlabs PF10-03-P01) on separate kinematic mounts (Thorlabs KM100) and through an aspheric lens (Thorlabs C260TMD-B) into a 50~\textmu m multi-mode fiber (Thorlabs M42L01) that guides the light to a photodiode (Thorlabs DET025AFC/M) for detection. A 600~nm long pass filter (Thorlabs FEL0600) and 900~nm short pass filter (Thorlabs FES0900) are placed in this path to ensure that only the \nv PL is detected.

\subsection{Microwave setup}\label{sec_MW}
A block diagram of the electronics part of our setup is shown in Figure~\ref{system_arch}(a). The block labeled `Microwave Source \& Amp' is discussed in this section. First, a PC communicates with the MW source (SignalCore SC800) to set the MW frequency between 0-6 GHz. The MW signal is fed into an I/Q modulator (Texas Instruments TRF370417EVM). The `I', or `in-phase', input of the modulator controls the amplitude of a MW signal that has the same phase as the input MW signal. The `Q', or `quadrature', input controls the amplitude of a MW signal that is 90$^{\circ}$ phase shifted. The output of the modulator is the sum of these signals. For our experiments, it is sufficient to restrict the I and Q inputs to high/low signals (TTL high or low), which enable or disable the I and Q signal components. The modulator's output is therefore either a signal that is in-phase with the input MW signal (I high, Q low), 90$^{\circ}$ out phase (I low, Q high), or 45$^{\circ}$ out of phase (I high, Q high). The modulator's output is then amplified by a MW amplifier (ZQL-2700MLNW+) to a MW power of +24~dBm (251~mW), and transmitted to the PCB antenna [shown in Figure~\ref{system_arch}(c)] with via a coaxial cable with SMA connectors on both ends.   

The antenna is designed to produce the oscillating magnetic field $B_{1}$ at the position of the \nv centers. Figure~\ref{system_arch}(c) shows our antenna design with all geometrical parameters, adapted from Ref.~\onlinecite{Sasaki2016}. The antenna was designed and simulated in-house using \textit{Eagle CAD} and \textit{CST MW Studio}, respectively, and manufactured commercially by the PCB manufacturer \textit{Circuit Labs}. The antenna geometry is referred to as a loop-gap resonator, where the `loop' in our case is a through hole, near which we can have a strong oscillating $B_{1}$ field in the direction perpendicular to the plane of the PCB as shown in Figure~\ref{system_arch}(d). This hole also allows for both the excitation laser and the PL of the diamond sample to pass through the PCB. 

Simulations predict that the antenna has a resonance at 2.57~GHz, indicated by the minimum of the S$_{11}$ reflection coefficient, shown by the blue curve in Figure~\ref{system_arch}(e). In the manufactured PCB, we measure the resonant S$_{11}$ minimum to be at $\sim$2.49~GHz (red curve), which agrees well with our simulated value.\cite{AntennaRes} The simulations also predict that with a MW power of +24~dBm (251~mW), the antenna produces a maximum oscillating magnetic field of $B_1=306$~\textmu T (black curve), which would result in an electron spin Rabi frequency of $\Omega_{\rm R}=\tfrac{1}{2}\gamma_e B_1\approx 4.3$~MHz.\cite{OscRot} The diamond sample is placed as close to the top layer [Figure~\ref{system_arch}(c)] of the PCB as possible (using tape as adhesive), and we align the excitation laser's focal spot to the center of the hole, where we can expect the greatest homogeneity of $B_{1}$ in magnitude ($\norm{\nabla{B_{1}}}=0.13$~\textmu T/\textmu m at +24 dBm of excitation power). While the antenna response peaks at $\sim$2.49~GHz, the bandwidth of the antenna is broad enough to perform the spin resonance measurements at 2.87~GHz (i.e. at $B_{0}=0$~T), which will be discussed in Section~\ref{sec_odmr}.

\begin{figure*}[!hbt]
	\centering 
        \includegraphics[width=\textwidth,keepaspectratio]{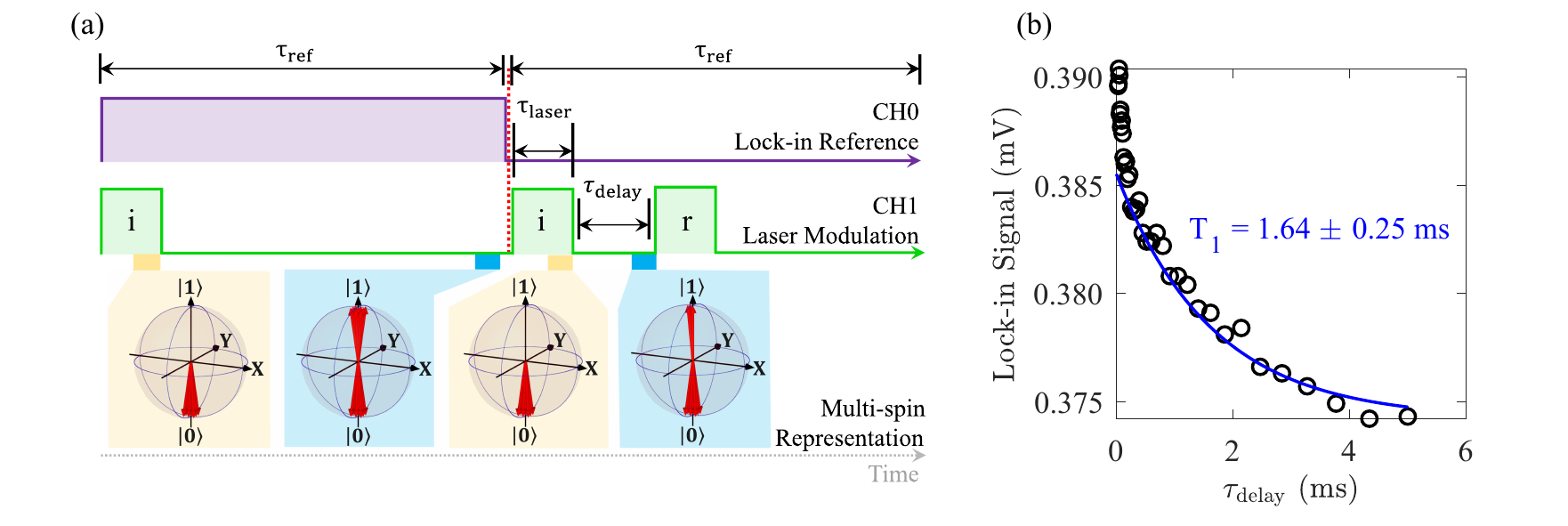}
        \caption{(a) Pulse sequences programmed to measure the spin $T_{1}$ decay of the \nv ground state, and a multi-spin representation showing the set of pure spin state vectors that constitute the \nv spin ensemble at selected times along the pulse sequence. $\tau_{\rm ref}$=15 ms, $\tau_{\rm laser}$=5~\textmu s, and $\tau_{\rm delay}$ is varied. (b) \nv center $T_{1}$ decay measured with lock-in detection. The total measurement time is $\sim$3 minutes.}
        \label{t1_experiment}
\end{figure*}

\subsection{Pulsing sequences and lock-in detection}\label{sec_LI}
As discussed in Section~\ref{sec_NV}, the \nv center emits ${\sim30}$\% more photons when decaying from the $\ket{0}_{\mathrm{e}}$ states compared to the $\ket{\pm1}_{\mathrm{e}}$ states. Under experimental conditions, where we collect emission from the four different \nv orientations simultaneously, and spin control of the ensemble is far from ideal, the contrast of the spin signal is limited to a few percent. It is possible to extract the spin signal by using a power-stabilized laser source and sufficient averaging. However, to achieve a more robust implementation for a teaching lab, we employ a lock-in amplifier. 

The operation of a phase-sensitive lock-in amplifier is well-known, and described in Ref.~\onlinecite{Scofield1994}. Given a reference frequency, the lock-in amplifier will make a phase sensitive measurement of signals present exactly at that frequency, with a bandwidth as narrow as 0.01 Hz (selectable). If the signal of choice can be modulated at the reference frequency, the lock-in amplifier can extract it with a good signal-to-noise ratio (SNR) from large background signals at other frequencies. For example, in an experiment where laser excitation provides a PL signal from an ensemble of \nv centers, and a resonant MW source drives the spin from the $\ket{0}_{\mathrm{g}}$ state to the $\ket{\pm1}_{\mathrm{g}}$ states, the spin signal can be extracted with a lock-in amplifier when the MW field is amplitude or frequency modulated at the reference frequency, while any DC signal is rejected. 

The experiments are clocked and triggered by a Pulse Blaster ESR Pro 250 pulse pattern generator, that can be programmed to generate TTL pulse sequences on up to 24 channels. For the experiments described here, we require 4 channels: 
\begin{itemize}
    \item `CH0' provides the reference frequency to a phase sensitive lock-in amplifier (Ametek 5210 or alternatively Stanford Research Systems SR830).
    \item `CH1' pulse-modulates the 520 nm laser diode via an in-house designed high-speed current source (discussed in Appendix~\hyperref[AppB]{B}).
    \item `CH2' provides the I input for the I/Q modulator (discussed in Section~\ref{sec_MW}).
    \item `CH3' provides the Q input for the I/Q modulator.
\end{itemize}

We perform our experiments by programming pulse sequences onto the 4 channels, keeping in mind that we only modulate the signals at the lock-in reference frequency that we are interested in detecting (except for the $T_1$ measurements in Section~\ref{sec_t1}). Each experiment below has a specific pulse sequence associated with it. The software programming of each pulse sequence is done on a PC using Matlab, and the pulse blaster is configured to generate the corresponding TTL pulses via USB serial interface.

\section{Experiments}\label{sec_Ex}

\subsection{Spin initialization and readout - \texorpdfstring{$T_1$}{T1} measurement}\label{sec_t1}
At room temperature, without any laser, MW excitation, or external magnetic fields, the \nv spins are completely depolarized and the spin states are very nearly equally populated, as described by a Boltzmann distribution for a two-level system with a 2.87~GHz level splitting at 300~K. Thus, all experiments discussed in this paper will start with spin initialisation via an optical pulse as described above. As discussed in Section~\ref{sec_NV}, electrons decaying via the $\ket{\mathrm{s}}$ state will be initialized into the $\ket{0}_{\mathrm{g}}$ state with a high probability. Hence, continuous 520 nm laser excitation, and therefore repeated optical cycling of the dynamics discussed in Section~\ref{sec_NV}, will result in the ensemble being initialized into the $\ket{0}_{\mathrm{g}}$ state.

Readout of the \nv spin states at the end of the experiment relies on the same process. A laser pulse is used to excite the \nv centers to the $\ket{\rm e}$ state. As relaxation via the $\ket{\rm s}$ state is more likely for the $\ket{\pm 1}_{\rm e}$ state than for the $\ket{0}_{\rm e}$ state, the $\ket{0}_{\mathrm{e}}$ state will result in a larger number of photons being detected at the wavelengths of the ZPL and PSB.

The spin $T_1$ decay time gives an idea of how long the spins remain in the prepared state, before longitudinal relaxation into a Boltzmann-distributed population occurs. Note, that a $T_1$ experiment measures the spin decay time only, and is not sensitive to any dephasing of the system. There are two methods to measure the spin $T_1$ decay time: \begin{enumerate}
    \item By initializing the spins into the $\ket{0}_{\rm g}$ state using a laser pulse, leaving them to relax for a time $\tau_{\rm delay}$, and then measuring the resultant spin population with a second laser pulse. This is a fairly simple measurement that requires no MW control and works at zero $B_{0}$ magnetic field. Furthermore, it results in a signal from all members of the ensemble, irrespective of the orientation of the \nv axes.
    \item By additionally using a resonant MW pulse to coherently invert the spin population before the decay period $\tau_{\rm delay}$. At non-zero $B_0$ field, this method allows the selection of a specific spin transition and measurement of its corresponding $T_1$ time. However, due to the different orientations of the \nv center axes with respect to the $B_0$ and $B_1$ fields (see also Section~\ref{sec_NV}), a MW pulse will not be resonant with all the defects or lead to non-perfect inversion, producing a partial initialization of the spin ensemble and resulting in a smaller signal.
\end{enumerate} 

We choose the first method for the teaching labs. This method results in a stronger signal, and allows the introduction of spin initialization and readout independent of the concept of MW spin control.

Figure~\ref{t1_experiment}(a) describes the pulse sequence to implement the all-optical $T_{1}$ measurement. CH0 sets the reference signal for the lock-in amplifier, while CH1 defines the laser pulse sequence. There are two initialization laser pulses in the sequence [denoted `i' in Figure~\ref{t1_experiment}(a)], one at the beginning of each half-cycle. As they are separated by a $\pi$-phase shift with respect to the lock-in reference (CH0), the lock-in detection will cancel out this signal under the condition that $\tau_{\rm ref}\gg T_1$. The second initialization pulse is followed by a readout laser pulse (denoted `r') of the same length, after a variable delay $\tau_{\rm delay}$. The signal caused by the readout pulse is dependent on the $T_{1}$ process, however since this laser pulse is only present in the second half-cycle of the reference, the readout pulse will additionally result in the detection of a PL signal that is independent of $T_{1}$ (i.e. a background signal). In Figure~\ref{t1_experiment}(b), we show the corresponding measurement. The signal decreases from 0.391~mV to 0.374~mV with a time constant of $T_1=1.64\pm0.25$~ms. The 4.3\% change in signal corresponds to the spin decay from the $\ket{0}_{\rm g}$ state to the Boltzmann-distributed population. The 0.374~mV offset originates from the spin-independent PL signal caused by the readout pulse being present in only one of the two half-cycles.

\begin{figure*}[hbt]
	\centering 
		\includegraphics[width=\textwidth]{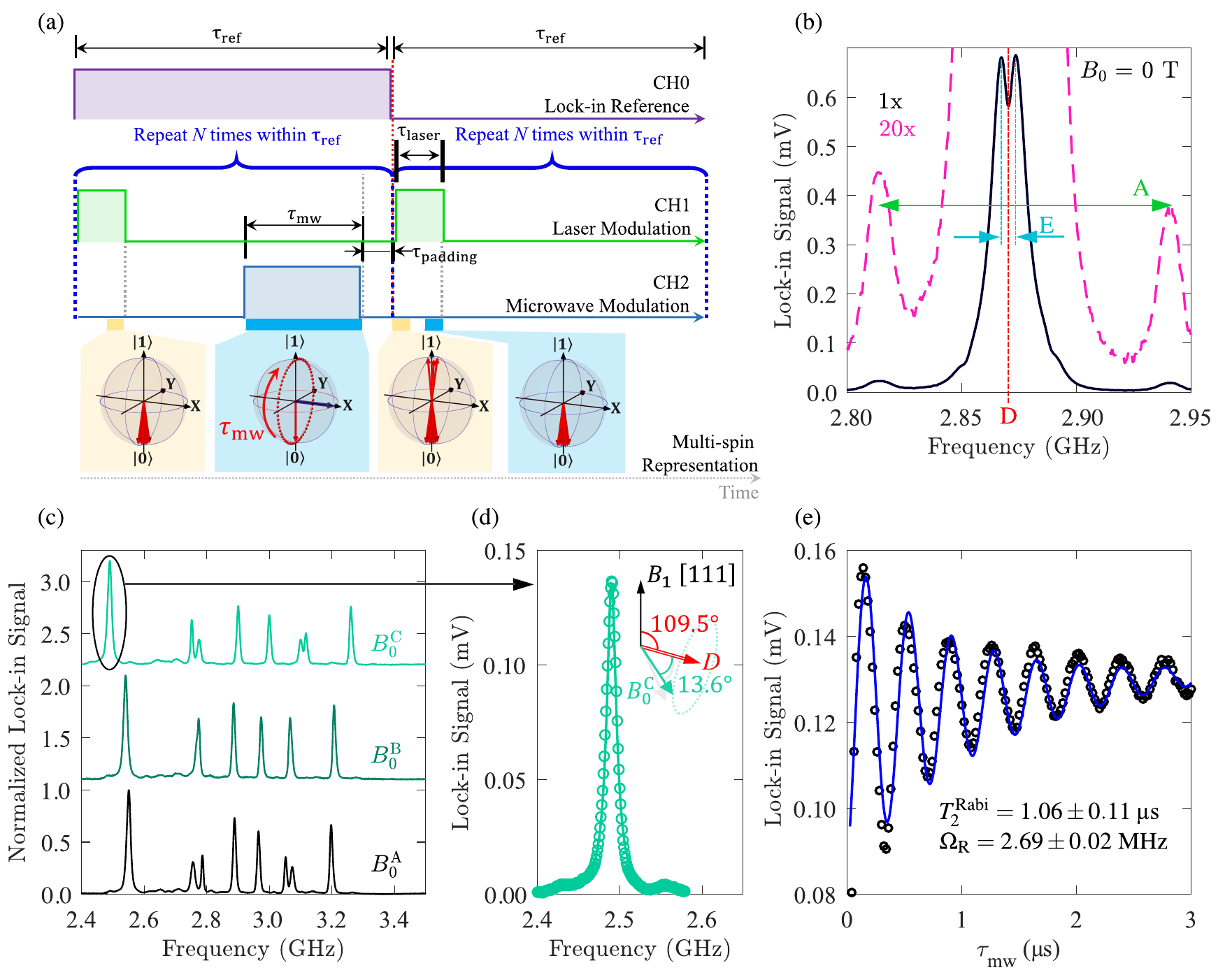}
		\caption{(a) Pulse sequence to perform pulsed-ODMR spectroscopy and Rabi oscillations, and multi-spin representation showing the set of pure spin state vectors on a Bloch sphere for \nv centers of one particular orientation (i.e. all spins are resonant with the MW frequency). $\tau_{\rm ref}=2.5$~ms, $\tau_{\rm laser}=5$~\textmu s, $\tau_{\rm padding}=1$~\textmu s, $\tau_{\rm mw}=5$~\textmu s for pulsed-ODMR, and $\tau_{\rm mw}$ is varied for Rabi oscillations. Laser and MW pulses sequences (CH1 and CH2) are repeated $N=250$ times within each half-cycle to increase the signal strength. (b) Zero-field pulsed-ODMR spectrum. The main ODMR peak is centered at $D=2.87$~GHz. The splitting $E=7$~MHz is a result of lifting the $\ket{\pm1}_{\mathrm{g}}$ degeneracy due to a spin-strain interaction (corresponding term not included in Eq.~\ref{eqn:hamiltonian}). The smaller side peaks are split by $A=128$~MHz, and are due to the hyperfine interaction of the \nv centers coupled to a nearest neighbor C$^{13}$ nucleus with a nuclear spin of $\pm{1/2}$ (corresponding term not included in Eq.~\ref{eqn:hamiltonian}). The total measurement time is $\sim$2 minutes. (c) Pulsed-ODMR signal obtained with a permanent magnet placed in 3 arbitrary locations within the vicinity of the sample. The $B_{0}$ magnitudes follow $B_{0}^{\rm A}<B_{0}^{\rm B}<B_{0}^{\rm C}$. The total measurement time is $\sim$6 minutes for each magnet position. (d) Peak from (c) for $B_{0}^{\rm C}$ at the resonant frequency of the MW antenna [compare Figure~\ref{system_arch}(d)]. The peak fits to a Lorentzian centered at $\nu_{\rm ODMR}=2.49$~GHz and linewidth $\Gamma_{\rm FWHM}=14.2$~MHz. (e) Rabi oscillations with MW frequency set to $\nu_{\rm ODMR}=2.49$ GHz. Measurement time is $\sim$8 minutes.}
		\label{odmr_rabi}
\end{figure*}

\begin{figure*}[hbt]
	\centering 
        \includegraphics[width=\textwidth]{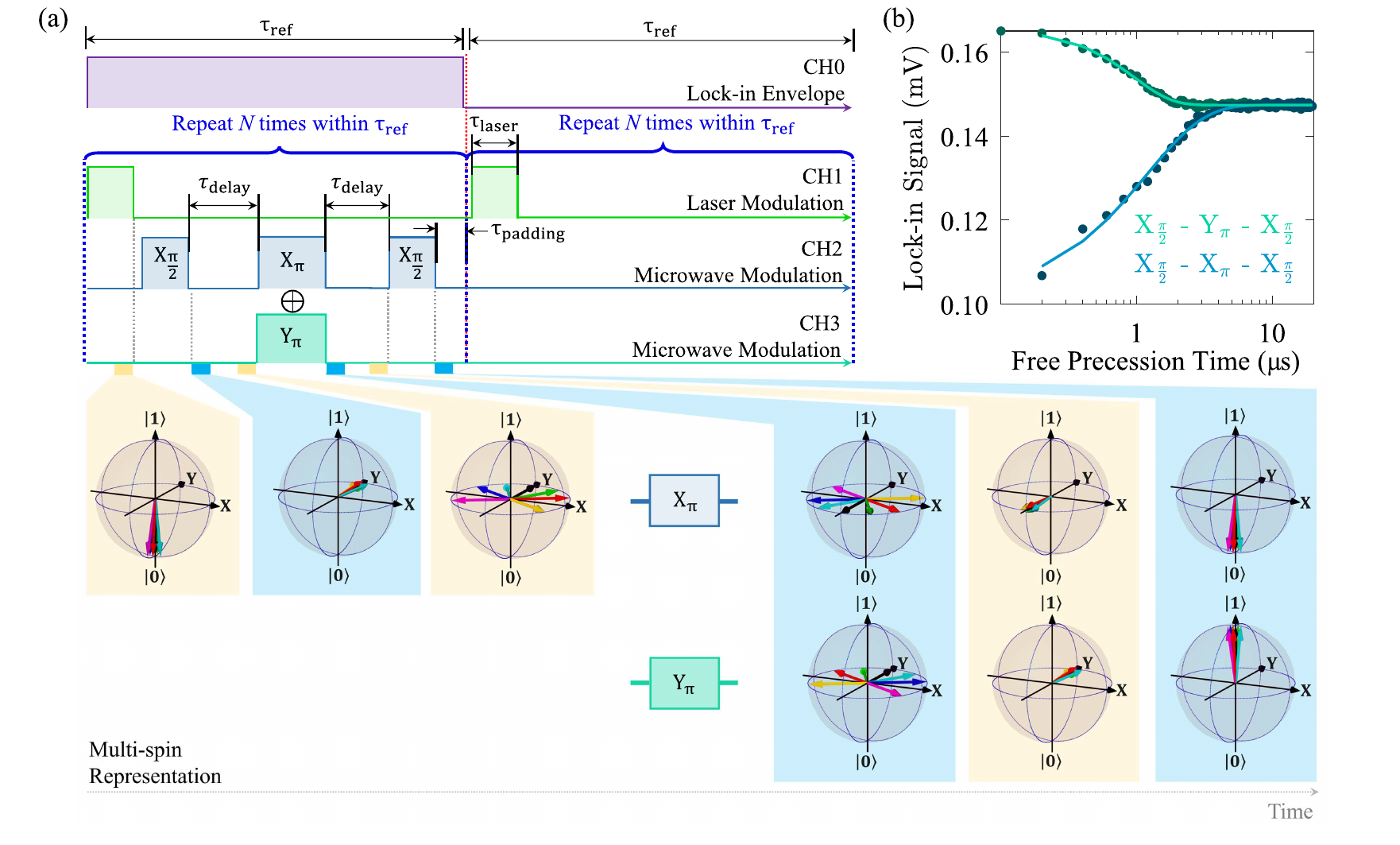}
        \caption{(a) Dynamical decoupling pulse sequence, with one refocusing pulse, where either a $\mathrm{X_{\pi}}$-pulse, or $\mathrm{Y_{\pi}}$-pulse can be chosen as the refocusing pulse. $\tau_{\rm delay}$ is varied, and the experiment is performed under the same experimental conditions as in Figure~\ref{odmr_rabi}(e), with $\tau_{\pi/2}=72$~ns and $\tau_\pi=144$~ns. Laser and MW pulses sequences (CH1, CH2, CH3) are repeated $N=100$ times within each half-cycle to increase the signal strength. The multi-spin representation shows the evolution of a set of pure spin state vectors on the Bloch sphere throughout the sequence.
        (b) Coherence time measurement for the echo sequence as in (a) with the MW $\pi$-pulse applied along the +X (Hahn echo) or +Y axis (1-pulse CPMG). For long free precession times, the ensemble enters a mixed state, as can be seen from the convergence of the two spin signals to a common value. The solid lines are fits to an exponential decay. Measurement time is $\sim$4 minutes for each scan.}
        \label{hahn_echo_experiment}
\end{figure*}

\subsection{Optically detected magnetic resonance and Zeeman effect}\label{sec_odmr}
As discussed in Section~\ref{sec_NV}, an oscillating magnetic field $B_{1}$ can be used to controllably rotate the \nv ground-state electronic spin. To demonstrate this, we employ another channel of the Pulse Blaster (denoted `CH2') to pulse-modulate the output of the MW source and create a sequence of oscillating $B_{1}$ pulses, as shown in Figure~\ref{odmr_rabi}(a) (see also Sections~\ref{sec_MW} and \ref{sec_LI} for details of the setup). We start by measuring the spin transition spectrum of the \nv center ensemble in zero magnetic field. As before, an initial laser pulse is used to initialize the \nv into the $\ket{0}$ state. A subsequent MW pulse of fixed length $\tau_{\rm mw}$ ($\gg T_{2}^{*}$) then attempts to rotate the spins. When the MW frequency is equal to the $\ket{0}_{\mathrm{g}}\longleftrightarrow\ket{\pm1}_{\mathrm{g}}$ transition frequency, we rotate the spins about the +X axis.\cite{ConvX} The next laser pulse will serve as the readout pulse. This procedure is referred to as optically detected magnetic resonance (ODMR), as the optical signal is reduced in magnitude when the spins are rotated from the $\ket{0}_{\mathrm{g}}$ state into the $\ket{\pm1}_{\mathrm{g}}$ state. Using a lock-in amplifier, the resonance condition will result in a large, positive magnitude reading, which represents the absolute value of the change in signal under resonance. Out of resonance the MW pulses have no effect on the lock-in magnitude. Figure~\ref{odmr_rabi}(b) shows the zero magnetic field ODMR spectrum. The main peak is centered at $D=2.87$~GHz, with a spin-strain splitting $E=7$~MHz which lifts the degeneracy of the $\ket{0}_{\mathrm{g}}\longleftrightarrow\ket{\pm1}_{\mathrm{g}}$ transition. Additionally, the side peaks visible are a result of hyperfine interaction ($A=128$~MHz) of the subset of \nv spins that are also coupled to a nearest neighbor C$^{13}$ nucleus. The side peaks are $\sim3.3\%$ of the center peak intensity, which corresponds to the probability of finding a C$^{13}$ nucleus next to an \nv site. 

Figure~\ref{odmr_rabi}(c) shows further ODMR spectra recorded for arbitrary magnetic fields $B_{0}^{\rm A}$, $B_{0}^{\rm B}$, $B_{0}^{\rm C}$, introduced by placing a permanent magnet in the proximity of the diamond crystal. The energy splitting of the $\ket{\pm1}_{\mathrm{g}}$ state depends on the orientation of the $B_{0}$ field with respect to the \nv axis, allowing us to distinguish between the 4 different \nv orientations in our ODMR spectra, and resolving a total of 8 transitions for the field strengths $B_{0}^{\mathrm{A}}$ and $B_{0}^{\mathrm{C}}$, and 6 transitions in $B_{0}^{\mathrm{B}}$ due to an overlap of 2 pairs of transitions (see also Section~\ref{sec_NV}).

\subsection{Rabi oscillations}\label{sec_Rab}

The term ``\textit{Rabi oscillations}'' refers to the driven evolution of a two-level system that manifests as the circular movement of the system's state vector around the Bloch sphere.\cite{Humble2019} Successful demonstration of Rabi oscillations means that we have achieved coherent control - an important step in demonstrating the viability of any quantum system (see also Section~\ref{sec_learn}). 
Figure~\ref{odmr_rabi}(d) shows a detailed ODMR spectrum of the $\ket{0}_{\mathrm{g}}\longleftrightarrow\ket{-1}_{\mathrm{g}}$ transition for $B_0^{\rm C}$. For this particular field, we have roughly aligned the magnetic field from the permanent magnet such that its direction is parallel to one of the three [1$\bar{1}\bar{1}$], [$\bar{1}1\bar{1}$], [$\bar{1}\bar{1}$1] \nv center axes, and from matching the spectrum in Figure~\ref{odmr_rabi}(c) to theory, we can extract $B_0=14.2$~mT with an angle of $13.6^\circ$ to the best-aligned \nv axis (see Figure~\ref{odmr_rabi}(d) inset). Furthermore, the position of the permanent magnet was adjusted such that the $\ket{0}_{\mathrm{g}}\longleftrightarrow\ket{-1}_{\mathrm{g}}$ ODMR transition frequency coincides with the resonant frequency of the PCB antenna (see Section~\ref{sec_MW}). This way, we subject our \nv spins to the largest $B_{1}$ field and hence achieve the fastest Rabi oscillation frequencies.\cite{B1Dir} This is important as short Rabi periods compared to the driven coherence time ensure a greater number of clearly observable oscillations, allowing us to perform the multi-pulse sequences described in Section~\ref{sec_DD}. A larger $B_{1}$ field will additionally result in an ODMR peak with the largest peak intensity and broadest linewidth.

To observe Rabi oscillations, we set the driving frequency $\nu_{\rm mw}$ to be in resonance with the ODMR transition frequency $\nu_{\rm ODMR}=2.49$~GHz, and vary the MW pulse length $\tau_{\rm mw}$ while recording the lock-in signal. We observe oscillations in the spin signal as a function of pulse length, as shown in Figure~\ref{odmr_rabi}(e), indicating coherent rotations of the spin around the Bloch sphere. As we have selected only one of the 8 visible transitions in Figure~\ref{odmr_rabi}(c), we are only coherently driving \nv centers of a single axis orientation and only from the $\ket{0}_{\mathrm{g}}$ to the $\ket{-1}_{\mathrm{g}}$ state. The oscillations can be fitted to an exponentially decaying sinusoid:
\begin{equation}
    \label{eqn:rabi_fit}
    V_{\rm LI}=A\cdot \mathrm{sin}(2\pi\Omega_{\mathrm R}\tau_{\mathrm {mw}}+\phi)\cdot\mathrm{e}^{-\tfrac{\tau_{\mathrm {mw}}}{T_2^{\mathrm{Rabi}}}}+B\tau_{\mathrm {mw}}+C,
\end{equation}
where $A$ is the Rabi oscillation amplitude, $B$ is a linear term included due to an observed increase in signal with increasing $\tau_{\mathrm{mw}}$ (possibly due to heating of the sample due at long $\tau_{\mathrm{mw}}$), $C$ is an offset, $\Omega_{\rm R}=2.69\pm0.02$~MHz is the Rabi frequency, $\phi=-1.24\pm0.09$ radians is the phase-offset ($-\frac{\mathrm{\pi}}{2}$ for ideal Rabi oscillations), and $T_2^{\rm Rabi}=1.12\pm0.14$~\textmu s is the driven coherence time of the spin.\cite{Yan2013,Laucht2017} The $T_2^{\rm{Rabi}}$ coherence time describes how long the $\ket{0}_{\mathrm{g}}\longleftrightarrow\ket{-1}_{\mathrm{g}}$ transition in the ensemble can be driven before the ensemble dephases into a mixed-state. This is due to both the inhomogeneity of the $B_{1}$ field over the region of the sample in focus, and the inhomogeneous broadening of resonance frequencies over the \nv centers being measured.

From the Rabi oscillations, we can calibrate the exact $\pi/2$- and $\pi$-pulse lengths as $\tau_{\pi/2}=72$~ns and $\tau_\pi=144$~ns. These pulse times will be important for the experiments in the following section (Section~\ref{sec_DD}), where we construct dynamical decoupling pulse sequences out of $\pi/2$- and $\pi$-pulses.

\subsection{Coherence times and dynamical decoupling}\label{sec_DD}

In the next experiment we measure the coherence time $T_2$ of the \nv centers. $T_2$ is the time over which a well-defined phase relation between a quantum state and a reference clock can be preserved, before noise and coupling to the environment randomize it. Compared to quantum systems operating at colder temperatures, room temperature systems are subject to a large amount of thermal energy that leads to a particularly noisy environment. As it is much easier to contemplate practical room temperature quantum systems than those that require complex cooling mechanisms, it becomes especially interesting to investigate their room temperature coherence times. $T_2$ is also one of the key metrics for comparing different quantum systems, however, as there are different definitions of $T_{2}$, it important to use the same metric when comparing different quantum systems. One such coherence time was already determined in Figure~\ref{odmr_rabi}(e) and is the coherence time of the system while it is driven ($T^{\rm{Rabi}}_{2}$). In the following experiments we look at the coherence time during free precession of the spins using different dynamical decoupling sequences.

Dynamical decoupling methods make use of refocusing pulses, as in the Hahn-echo sequence~\cite{Hahn1950} [see also Figure~\ref{hahn_echo_experiment}(a)], to refocus the phases of spins that precess at slightly different rates. Here, one differentiates between inhomogeneous dephasing where the transition frequencies of individual \nv centers are shifted due to their local environments, e.g. due to static inhomogeneities in the sample itself or in the applied $B_0$ magnetic field, and homogeneous dephasing where the transition frequencies of all \nv centers are broadened by similar amounts, e.g. due to dynamic noise or the finite lifetime of the quantum state. Sequences consisting of refocusing pulses are very good at refocusing static shifts in transition frequencies -- a single refocusing pulse (as in the Hahn echo) is sufficient to decouple the quantum system from static noise.\cite{Hahn1950} Noise of any frequency is refocused as long as its frequency is much lower or much higher than the refocusing pulse repetition frequency, which can be best understood in the filter function formalism as described in Refs.~\onlinecite{Biercuk2011,Bylander2011}. In fact, changing the pulse repetition frequency changes the frequency spectrum the spin remains sensitive to, which allows conducting detailed investigations of the noise spectrum.\cite{Alvarez2011,Bylander2011,Muhonen2014,Chan2018} 

In Figure~\ref{hahn_echo_experiment}(a) we show the pulse sequence used to measure the Hahn echo coherence time $T_2^{\rm{Hahn}}$, while the Bloch spheres at the bottom of the panel give an indication of the spin orientations at specific points of the sequence. A first X$_{\pi/2}$-pulse rotates the spin to the +Y direction on the Bloch sphere. The spins are left to freely precess for a time $\tau_{\rm delay}$, before a X$_\pi$-pulse rotates them to the -Y direction. Any phase that they might have accumulated with respect to the rotating frame until then (as indicated by the coloured arrows) will be unwound in the second free precession time $\tau_{\rm delay}$, before a final X$_{\pi/2}$-pulse rotates them to the +Z direction for readout. We present the corresponding experimental data in Figure~\ref{hahn_echo_experiment}(b). Here, $\tau_{\rm delay}$ is increased until any phase relation is randomized and the spin signal saturates, indicating a completely mixed state. The spin refocusing pulse can be applied along either the +X (Hahn echo) or +Y axis (using IQ modulation as described in Section~\ref{sec_MW}), with the system entering into the same mixed-state in either case as shown in Figure~\ref{hahn_echo_experiment}(b). We fit the data to an exponential decay and extract a coherence time of $T_2^{\rm Hahn}=1.2\pm0.2$~\textmu s.

\begin{figure}[!t]
	\centering 
		\includegraphics[width=\columnwidth]{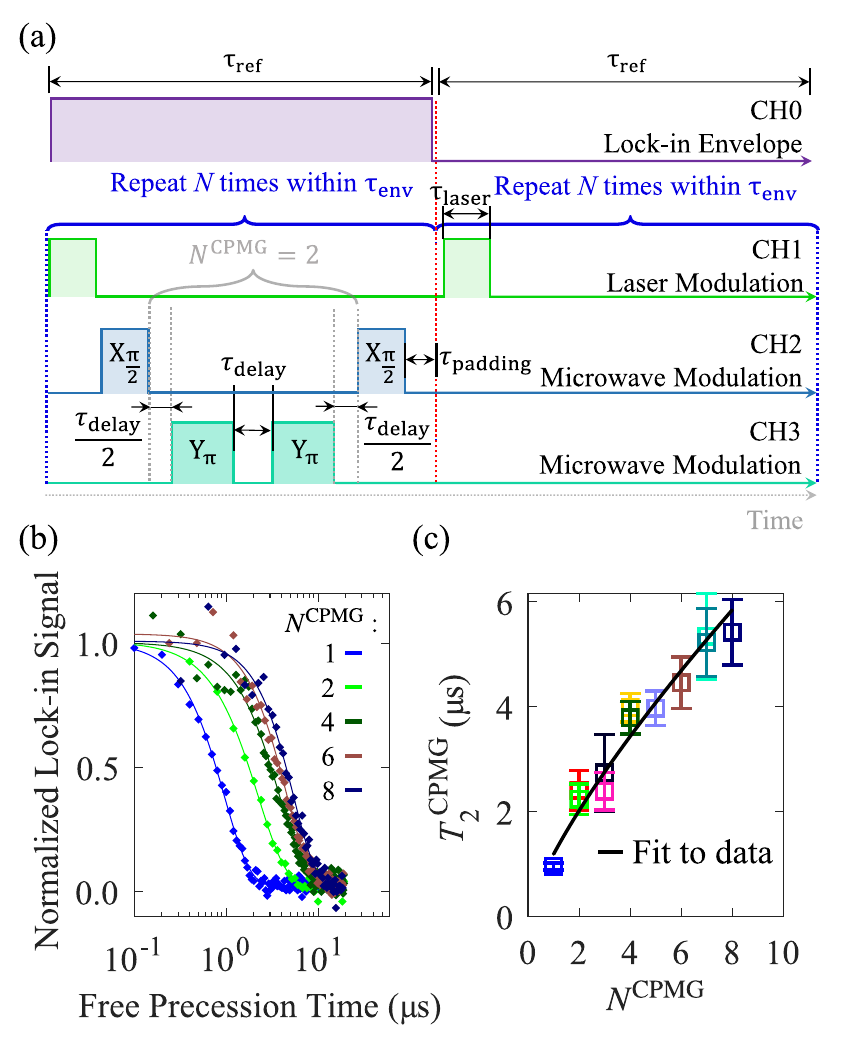}
		\caption{
		(a) Pulse sequences for the CPMG spin-echo sequence, where $N^{\rm{CPMG}}$ is number of Y$_{\pi}$ refocusing pulses. Laser and MW pulse sequences (CH1, CH2, CH3) are repeated $N=100$ times within each half-cycle to increase the signal strength.
		(b) Examples of CPMG measurements for different $N^{\rm{CPMG}}$. The measurement time is $\sim$4 minutes for each scan. 
		(c) Extracted $T_{2}^{\rm CPMG}$ for all CPMG scans.
		}
		\label{cpmg_experiment}
\end{figure}

Instead of a single refocusing pulse, multiple such refocusing pulses can be applied to further extend the spin coherence time. However, performing multiple refocusing pulses about the +X axis leads to an accumulation of pulse errors. Hence, refocusing pulses about the +Y axis -- known as Carr-Purcell-Meiboom-Gill (CPMG) spin-echo pulse sequence~\cite{Carr1954,Meiboom1958} -- are more advantageous. For the CPMG pulse sequence shown in Figure~\ref{cpmg_experiment}(a), $N^{\rm{CPMG}}$ is the total number of Y$_{\pi}$-pulses, and $t=N^{\rm{CPMG}}\tau_{\rm delay}$ is the total free precession time of the \nv spin. The CPMG pulse sequence acts as a bandpass filter -- increasing the number of refocusing pulses for a fixed $\tau_{\rm delay}$ sharpens the filter, whereas decreasing the time $\tau_{\rm delay}$ between refocusing pulses has the effect of shifting the center of the filter to higher frequency. The bandpass center frequency is given by $\frac{\pi}{\tau_{\rm delay}}$.\cite{Bylander2011} 

In Figure~\ref{cpmg_experiment}(b), we plot the result of CPMG sequences with an increasing number of refocusing pulses $N^{\rm{CPMG}}$. For the same total free precession time, a larger $N^{\rm{CPMG}}$ implies a shorter $\tau_{\rm delay}$, and hence a larger center frequency of the CPMG  filter function.\cite{Bar-Gill2012} The normalized lock-in signal as a function of total free precession time can be fitted to $C(t)=A\mathrm{exp}[(-\frac{t}{T_{2}^{\rm CPMG}})^n]$. In Figure~\ref{cpmg_experiment}(c), we plot the extracted $T_{2}^{\rm CPMG}$ as a function of $N^{\mathrm{CPMG}}$ and observe a clear correlation between $T_{2}^{\rm CPMG}(N^{\rm{CPMG}})$ and $N^{\rm{CPMG}}$, suggesting that the noise spectral density of the \nv electron's environment reduces towards higher frequencies. We were able to fit this data with a function of the form $T_{2}^{\rm CPMG}(N^{\rm{CPMG}})=B(N^{\rm{CPMG}})^{\alpha}$ with $\alpha=0.77 \pm 0.13$, which is within the bounds experimentally determined in Ref.~\onlinecite{Bar-Gill2012} for a CVD grown diamond with a large \nv density ($\sim10^{16}\rm{cm}^{-3}$).

\section{Conclusion}
We have presented a cost-effective experimental setup that is suitable for demonstrating coherent spin control concepts in an undergraduate teaching laboratory environment. The experiments implement the optically detected magnetic resonance technique at room temperature on \nv centers in diamond, and the setup is constructed using the minimal necessary optics and electronics components. Students will develop an intuitive feeling for quantum spin physics, gain first-hand experience in controlling a quantum system, and have the freedom to develop unique pulse sequences and observe the results in real-time. The use of a high-density \nv diamond sample provides a large signal, making the measurements insensitive to misalignment of the optics and exposure to high ambient light levels -- as is desirable for an undergraduate lab setup.\\

\clearpage

\onecolumngrid
\appendix

\section*{Appendix A: List of parts}\label{AppA}

\pgfplotstableset{
    highlightrow/.style={
        postproc cell content/.append code={
           \count0=\pgfplotstablerow
            \ifnum\count0=#1
               \pgfkeysalso{@cell content=\textbf{##1}}
            \fi
        },
    },
}

\begin{table}[htpb]
    \centering
    \renewcommand{\arraystretch}{0.99}
    \resizebox{\textwidth}{!}{
    \begin{filecontents*}{parts_optics.csv}
Item,Supplier,Part Number,Application,Price (USD),Qty,Total (USD)
,,,,(17/12/2019),,
Housing,,,,,,
Aluminum Breadboard,Thorlabs,MB3045/M,Base plate,199.45,1,199.45
Enclosure,Thorlabs,XE25C7/M,Enclosure for all optics,199.11,1,199.11
Optical Excitation,,,,,,
SM Fiber-Pigtailed Laser Diode,Thorlabs,LP520-SF15,Excitation laser,717.45,1,717.45
Laser Diode ESD Protection,Thorlabs,SR9HA ,Laser diode protection,54.11,1,54.11
Fiber Adapter Plates,Thorlabs,SM1FC - FC/PC ,Laser fiber coupling,31.38,1,31.38
Kinematic Mount,Thorlabs,KM100T,Laser mounting,70.87,1,70.87
50 mm Pedestal Pillar Post,Thorlabs,RS2P/M,Laser mounting,28.95,1,28.95
Aspheric Lens,Thorlabs,C280TMD-A ,Laser collimation lens,85.49,1,85.49
SM1 to M9 Lens Cell Adapter,Thorlabs,S1TM09,Lens mounting,24.35,1,24.35
SM1 Lens Tube 1.50in,Thorlabs,SM1L15,Lens mounting,16.17,1,16.17
Longpass 550 nm dichroic mirror,Thorlabs,DMLP550 ,Excitation/detection filtering,182.88,1,182.88
Kinematic Mount,Thorlabs,KM100,Dichroic mirror mount,39.86,1,39.86
50 mm Pedestal Pillar Post,Thorlabs,RS2P/M,Dichroic mirror mount,28.95,1,28.95
Microscope Objective,Olympus ,MS Plan 50x/0.80NA,Excitation lens,744.00,1,744.00
RMS Threaded Cage Plate,Thorlabs,CP42/M,Objective lens mount,32.78,1,32.78
50 mm Pedestal Pillar Post,Thorlabs,RS2P/M,Objective lens mount,28.95,1,28.95
5 mm Post Spacer,Thorlabs,RS5M,Objective lens mount,8.17,1,8.17
Sample Mount,,,,,,
3-axis Sample Positioner,Thorlabs,MBT616D/M,Stage for diamond sample,1184.92,1,1184.92
Blank Device Mount,Thorlabs,HBB002,Sample mount,56.55,1,56.55
Imaging,,,,,,
Protected Silver Mirror,Thorlabs,PF10-03-P01,Imaging mirror,53.58,1,53.58
Flip Mount,Thorlabs,TRF90/M,Mirror mounting,88.73,1,88.73
38 mm Pedestal Pillar Post,Thorlabs,RS1.5P4M,Mirror mounting,25.21,1,25.21
600 nm Longpass Filter (optional),Thorlabs,FEL0600,Imaging filtering,80.62,1,80.62
900 nm Shortpass filter (optional),Thorlabs,FES0900,Imaging filtering,80.62,1,80.62
SM1 Lens Tube 0.50in (optional),Thorlabs,SM1L05,Filter mounting,12.97,2,25.94
Achromatic Doublet,Thorlabs,AC254-125-A,Imaging lens,81.16,1,81.16
SM1 Lens Tube 1.50in,Thorlabs,SM1L15,Lens mounting,16.17,1,16.17
CMOS Camera,Thorlabs,DCC1545M ,Imaging camera,387.92,1,387.92
SM1 Lens Tube Spacer 3.50in,Thorlabs,SM1S35,Lens/camera mounting,25.53,1,25.53
SM1 Cage Plate,Thorlabs,CP33/M,Lens/camera mounting,16.89,2,33.78
50 mm Pedestal Pillar Post,Thorlabs,RS2P/M,Objective lens mount,28.95,2,57.90
5 mm Post Spacer,Thorlabs,RS5M,Objective lens mount,8.17,2,16.34
Detection,,,,,,
Protected Silver Mirror,Thorlabs,PF10-03-P01,Detection mirror,53.58,2,107.16
Kinematic Mount,Thorlabs,KM100,Mirror mounting,39.86,2,79.72
50 mm Pedestal Pillar Post,Thorlabs,RS2P/M,Mirror mounting,28.95,2,57.90
600 nm Longpass Filter,Thorlabs,FEL0600,Detection filtering,80.62,1,80.62
900 nm Shortpass filter,Thorlabs,FES0900,Detection filtering,80.62,1,80.62
SM1 Lens Tube 0.50in,Thorlabs,SM1L05,Filter mounting,12.97,2,25.94
Aspheric Lens,Thorlabs,C260TMD-B,Fiber coupling lens,85.49,1,85.49
SM1 to M9 Lens Cell Adapter,Thorlabs,S1TM09,Lens mount,24.35,1,24.35
XY-Axes Translation Mount,Thorlabs,ST1XY-D/M,Fiber alignment,517.26,1,517.26
38 mm Pedestal Pillar Post,Thorlabs,RS1.5P4M,Fiber coupler mounting,25.21,1,25.21
4 mm Post Spacer,Thorlabs,RS4M,Fiber coupler mounting,7.90,1,7.90
Z-Axis Translation Mount,Thorlabs,SM1Z,Fiber alignment,205.60,1,205.60
2-inch Rods,Thorlabs,ER2,Fiber alignment,6.28,4,25.12
Fiber Adapter Plates,Thorlabs,SM1FC - FC/PC ,Fiber coupling,31.38,1,31.38
50 micron FC/PC Multimode Fiber,Thorlabs,M42L01,Detection fiber,70.87,1,70.87
Si Photodetector,Thorlabs,DET025AFC/M ,Signal detection,306.00,1,306.00
Mounting,,,,,,
M6 Clamping Fork,Thorlabs,CF125C/M,Mounting,11.69,9,105.21
Sample,,,,,,
CVD Diamond Sample ,Element Six,145-500-0274-01,Sample,130.00,1,130.00
(e.g. SC Plate CVD $\langle$100$\rangle$),,,,,,
B0 and B1 Magnetic Fields,,,,,,
PCB Antennas (pack of 5), Circuit Labs,,B1 Microwave Excitation,72.16,1,72.16
Neodymium Block Magnets 10x10x5mm, AMF Magnets,,B0 Static Magnetic Field,2.30,4,9.20
Total Optics Cost:,,,,,,6780.60
\end{filecontents*}
\pgfplotstabletypeset[
    col sep=comma,
    assign column name/.style={/pgfplots/table/column name={\textbf{#1}}},
    every head row/.style={before row=\toprule,after row=\midrule},
    every row no 1/.style={before row=\hline,after row=\midrule},
    every row no 4/.style={before row=\hline,after row=\midrule},
    every row no 20/.style={before row=\hline,after row=\midrule},
    every row no 23/.style={before row=\hline,after row=\midrule},
    every row no 37/.style={before row=\hline,after row=\midrule},
    every row no 54/.style={before row=\hline,after row=\midrule},
    every row no 56/.style={before row=\hline,after row=\midrule},
    every row no 59/.style={before row=\hline,after row=\midrule},
    every last row/.style={before row=\hline,after row=\bottomrule},
    highlightrow={0},
    highlightrow={0},
    highlightrow={1},
    highlightrow={4},
    highlightrow={20},
    highlightrow={23},
    highlightrow={37},
    highlightrow={54},
    highlightrow={56},
    highlightrow={59},
    highlightrow={62},
    display columns/0/.style={string type,column type=l},
    display columns/1/.style={string type,column type=l},
    display columns/2/.style={string type,column type=l},
    display columns/3/.style={string type,column type=l},
    display columns/4/.style={string type,column type=r},
    display columns/5/.style={string type,column type=c},
    display columns/6/.style={string type,column type=r},
    ]{parts_optics.csv}
    }
    \caption{List of parts for `Optics' section of experimental set-up, shown in Figure~\ref{system_arch}(b).}
\end{table}

\clearpage
\newpage

\begin{table}[htpb]
    \centering
    \resizebox{\textwidth}{!}{
    \begin{filecontents*}{parts_electronics.csv}
Item,Supplier,Part Number,Application,Price (USD),Qty,Total (USD)
,,,,(17/12/2019),,
USB Pulse Blaster,SpinCore,PBESR-PRO-250-USB,Pulse Sequences,4485.00,1,4485.00
Dual Phase Lock-In,Ametek,5210,Lock-in detection,2000.00,1,2000.00
Windows PC,,,,2000.00,1,2000.00
6 GHz Microwave Source,SignalCore,SC800,Microwave Drive,1295.00,1,1295.00
Benchtop 3-Channel PSU,Newark,HM7042-5.02,Microwave Drive Power,1155.00,1,1155.00
RF Amplifier,Mini-Circuits,ZQL-2700MLNW+,Microwave Drive,304.95,1,304.95
IQ Modulator,Texas Instr.,TRF370417EVM,Microwave Drive,199.00,1,199.00
12V DC Wall Adapter,Thorlabs,LDS12B,Home-built current source,85.22,1,85.22
RS232 Serial Adapter,StarTech,ICUSB232DB25,Lock-in detection,20.99,1,20.99
BNC-BNC Coaxial,Thorlabs,2249-C-48,TTL Signals,19.29,5,96.45
SMA-SMA Coaxial 12-inch,Thorlabs,CA2912,Microwave interconnects,15.69,4,62.76
SMA 50-ohm termination,Mini-Circuits,ANNE-50X,IQ Modulator,15.95,2,31.90
BNC-SMA Adapter,Thorlabs,T4290,IQ Modulator,14.07,2,28.14
USB A-Mini B Cable,Thorlabs,USB-AB-72,Microwave Source,8.87,1,8.87
USB A-B Cable,Thorlabs,USB-A-79,Pulse Blaster,8.87,1,8.87
BNC-Banana Adapter,Thorlabs,T1452,Laser Diode ESD Protection,8.65,1,8.65
Banana Patch Cable,Thorlabs,T13120(2),Microwave Drive Power,7.31,4,29.24
Total Electronics Cost:,,,,,,11820.04
\end{filecontents*}
\pgfplotstabletypeset[
    col sep=comma,
    assign column name/.style={/pgfplots/table/column name={\textbf{#1}}},
    every head row/.style={before row=\toprule,after row=\midrule},
    every row no 1/.style={before row=\hline,after row=\midrule},
    every last row/.style={before row=\hline,after row=\bottomrule},
    highlightrow={0},
    highlightrow={0},
    highlightrow={18},
    display columns/0/.style={string type,column type=l},
    display columns/1/.style={string type,column type=l},
    display columns/2/.style={string type,column type=l},
    display columns/3/.style={string type,column type=l},
    display columns/4/.style={string type,column type=r},
    display columns/5/.style={string type,column type=c},
    display columns/6/.style={string type,column type=r},
    ]{parts_electronics.csv}
    }
    \caption{List of parts for `Electronics' section of experimental set-up, shown in Figure~\ref{system_arch}(a).}
\end{table}

\clearpage

\section*{Appendix B: Design of the laser driver}\label{AppB}
\begin{figure}[hbt]
	\centering 
		\includegraphics[width=\textwidth]{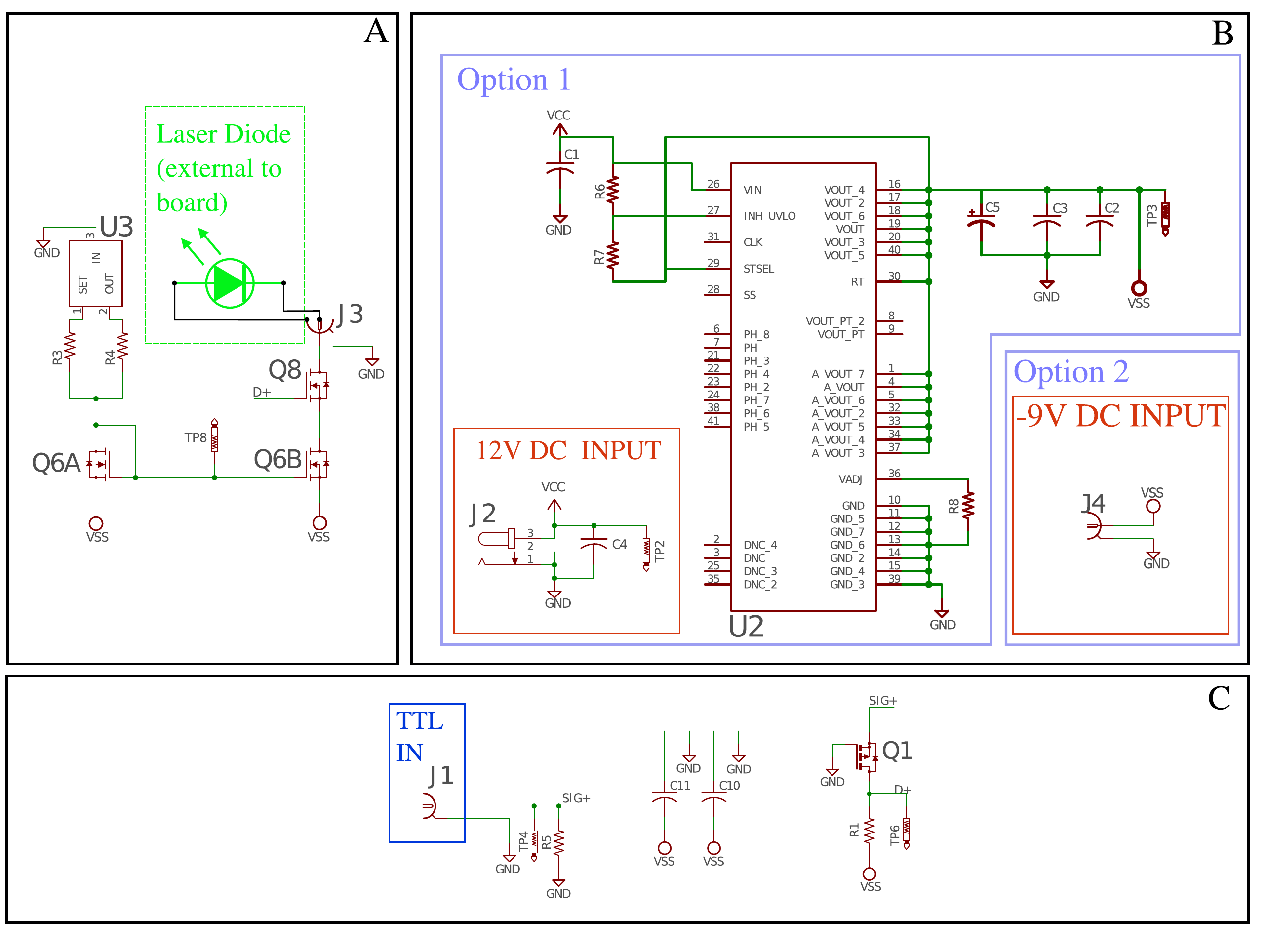}
		\caption{Schematic of the custom current source used in this experiment, broken down into the three main blocks. \textbf{A}. The circuitry supplying the current to the diode. U3 is an off-the-shelf current source integrated circuit that has the output current set by R3 and R4 to $\sim$165~mA. As U3 requires some time to settle to a steady state current, if we were to modulate this current directly, we would not be able to operate at the desired bandwidth of 10~MHz. Thus we construct a current mirror using Q6; a matched pair of N-channel MOSFETs. The mirror sources 165~mA through the diode, unless the connection is interrupted by Q8. This allows U3 to output a constant, stable current, circumventing the switching speed problems. \textbf{B}. Power supplies required for the operation of the circuit. There are two options available for the operation of this circuit. In `Option 1', an external 12~V input can be provided, which is then regulated down to Vss by U2. In `Option 2', a lab power supply can be directly connected to the circuit to provide the power. Vss was set to -9~V as a safety feature. The case of the laser diode is at the electrical potential of the anode and thus, this potential was set to ground. As a consequence, a negative supply voltage is required. For the current source that was used to obtain the results in this paper, there was a soldering error with the power supply provided by U2 in `Option 1', and thus `Option 2' was used. We have populated another PCB with `Option 1' for another teaching setup and confirmed that it functions correctly. \textbf{C}. Input logic and a level shifter. This block takes the TTL input signal and level-shifts it between ground and Vss. Such voltage levels are required to correctly drive the MOSFETs in block \textbf{A}, as standard TTL levels would not be able to switch Q8 due to the presence of negative supply voltages.}
		\label{current_source}
\end{figure}

\clearpage

\begin{table}
    \centering
    \resizebox{\textwidth}{!}{
    \begin{filecontents*}{parts_current_source_rev_2.csv}
Component,Description,Supplier,Part,Price (USD),Qty,Total (USD)
,,,Number,(17/12/2019),,
U1,LDO 3V3 Regulator,Element 14,1469102,1.6,1,1.6
U2,Integrated Power Supply,RS Online,798-1290,19.05,1,19.05
U3,Current Source,RS Online,779-9615,5.58,1,5.58
J1 J3 J4,BNC Connector,Element 14,1169739,7.25,3,21.75
J2,Barrel Jack,Element 14,1854514,1.25,1,1.25
Q1,PFET,Element 14,1510765,0.27,1,0.27
Q8,NFET,Element 14,2317616,0.2,1,0.2
Q6,Dual NFET,Element 14,2706719,1.07,1,1.07
R1,\SI{1}{\kilo\ohm} Resistor 0603,Element 14,2284191,0.14,1,0.14
R3,\SI{16.5}{\kilo\ohm} Resistor 0805,Element 14,2483960,0.46,1,0.46
R4,\SI{1}{\ohm} Resistor 0805,Element 14,2813642,0.31,1,0.31
R5,\SI{10}{\kilo\ohm} Resistor 0603,Element 14,1652827,0.19,1,0.19
R6,\SI{205}{\kilo\ohm} Resistor 0603,Element 14,2303251,0.01,1,0.01
R7,\SI{75}{\kilo\ohm} Resistor 0603,Element 14,1799354,0.01,1,0.01
R8,\SI{102}{\kilo\ohm} Resistor 0603,Element 14,2336927,0.21,1,0.21
C1 C4,\SI{10}{\micro\farad} Capacitor 1206 ,Element 14,2118134,0.61,2,1.22
C2 C3 C7,\SI{1}{\micro\farad} Capacitor 0603,Element 14,1845736,0.16,3,0.48
C6 C8 C9 C10 C11,\SI{100}{\nano\farad} Capacitor 0603,Element 14,1709958,0.02,5,0.1
C5,\SI{100}{\micro\farad} Capacitor 0603,Element 14,2354734,1.57,1,1.57
PCB,,Seeed Studio,,57.86,1,57.86
Total PCB Cost:,,,,,,113.33
\end{filecontents*}
\pgfplotstabletypeset[
    col sep=comma,
    assign column name/.style={/pgfplots/table/column name={\textbf{#1}}},
    every row no 1/.style={before row=\hline,after row=\midrule},
    every last row/.style={before row=\hline,after row=\bottomrule},
    highlightrow={0},
    highlightrow={0},
    highlightrow={21},
    highlightrow={21},
    display columns/0/.style={string type,column type=l},
    display columns/1/.style={string type,column type=l},
    display columns/2/.style={string type,column type=l},
    display columns/3/.style={string type,column type=l},
    display columns/4/.style={string type,column type=r},
    display columns/5/.style={string type,column type=c},
    display columns/6/.style={string type,column type=r},
    ]{parts_current_source_rev_2.csv}
    }
    \caption{List of parts required to assemble current driver circuit in Figure~\ref{current_source}.}
\end{table}

\begin{figure*}[hbt]
	\centering 
        \includegraphics[width=\textwidth]{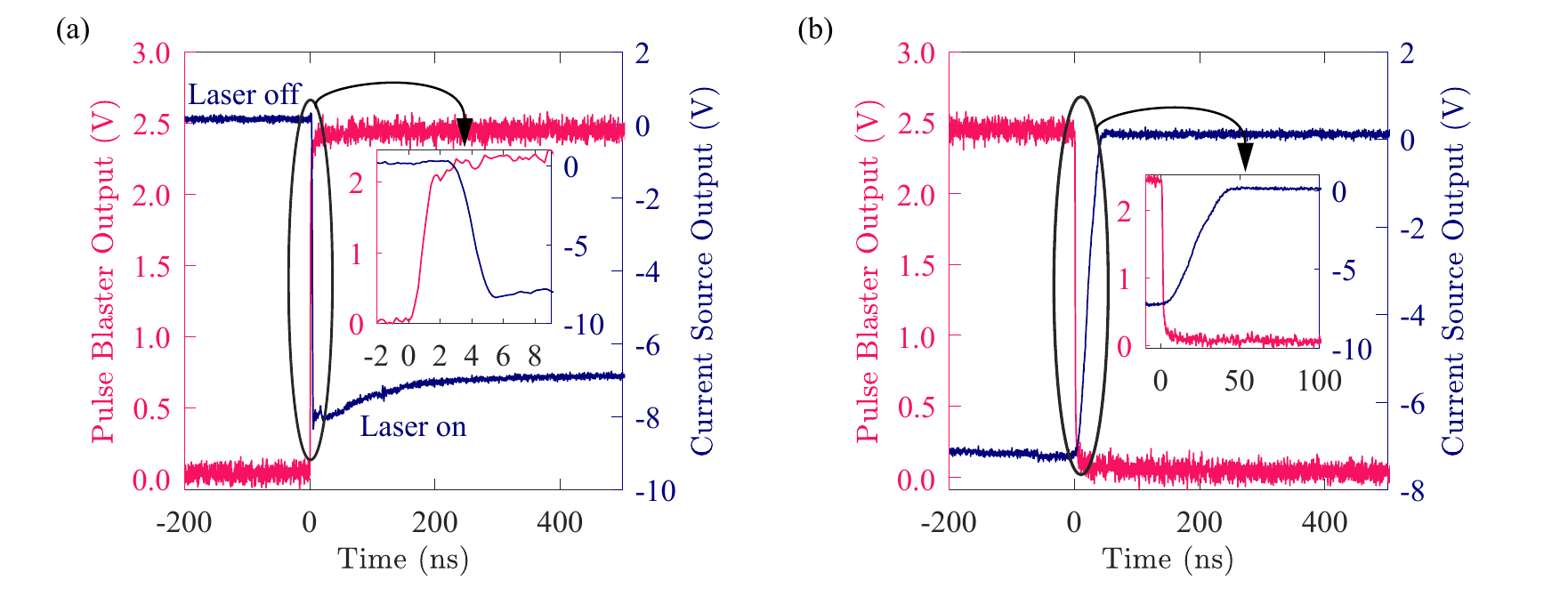}
        \caption{(a,b) Performance characterisation of the current driver, showing (a) turn-on and (b) turn-off of the laser. The pink curve is a copy of the pulse blaster TTL output that is used as an input for the current driver. The blue curve is the output of the current driver at its falling edge (a) and rising edge (b), respectively. The traces were measured with a Tektronix TDS3052 oscilloscope with 250 MHz bandwidth. The input impedances of both oscilloscope channels were 50~$\Omega$. The fall time of the current source output (turn-on time) is 1.6 ns and the rise time (turn-off time) is 27.0 ns.}
        \label{current_trace}
\end{figure*}

\clearpage

\section*{Appendix C: Comparison of HPHT and CVD diamond}\label{AppC}
\begin{figure*}[hbt]
	\centering 
        \includegraphics[width=\textwidth]{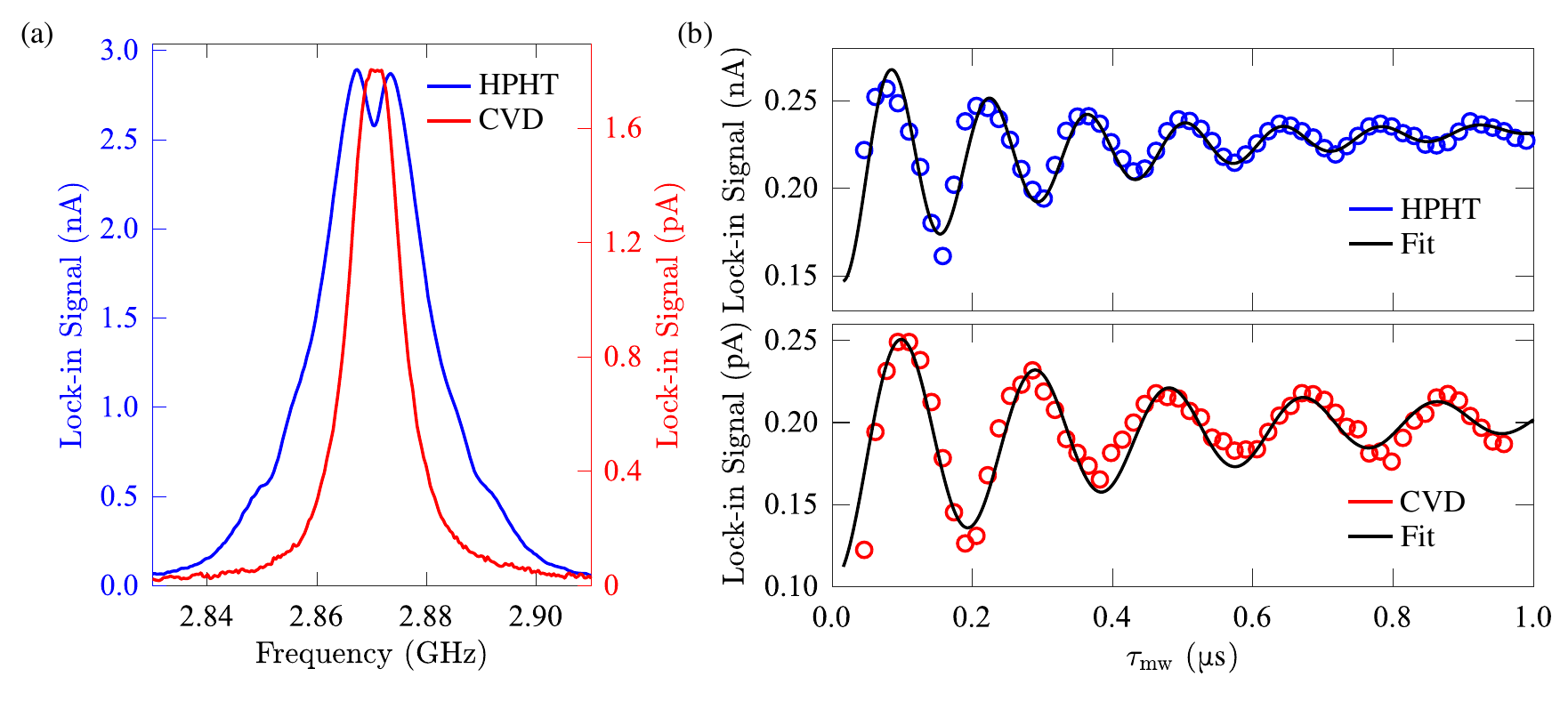}
        \caption{Measurements comparing the HPHT diamond used for all experiments in the main text (blue) and a CVD diamond (red) purchased from Element Six, listed in Appendix~\hyperref[AppA]{A}. Measurements were taken in current mode (similar signal-to-noise ratios were observed for voltage and current mode) of the EG\&G 5210 lock-in amplifier listed in Appendix~\hyperref[AppA]{A}. The PL emitted was 705~nW by the HPHT diamond and 0.75~nW by the CVD diamond, as measured with a power meter for $\sim15$~mW excitation power at $\lambda_{\rm{exc}}=520$~nm. 
        (a) Zero-field ODMR comparison. Measurement time was $\sim$2 minutes for the HPHT diamond and $\sim9$ minutes for the CVD diamond for a spectrum with 200 data points. 
        (b) Rabi oscillations comparison at $\nu_{\rm{ODMR}}=2.53$~GHz. Measurement time was $\sim2$ minutes for the HPHT diamond and $\sim13$ minutes for the CVD diamond for 60 data points. Fits were performed with Eq.~\ref{eqn:rabi_fit}, which gives $T_{2}^{\mathrm{Rabi}}=310\pm60$~ns and $\Omega_{\mathrm{R}}=7.2\pm0.1$~MHz for the HPHT diamond, and $T_{2}^{\mathrm{Rabi}}=450\pm90$~ns and $\Omega_{\mathrm{R}}=5.2\pm0.1$~MHz for the CVD diamond. 
        In these measurements, $T_{2}^{\mathrm{Rabi}}$ is limited by the inhomogeneous broadening of the ensemble linewidth, e.g. due to strain and hyperfine coupling to nuclear spins, and the inhomogeneity of the $B_{1}$ field.
        The PCB antenna used to perform these measurements had the same dimensions as described in Fig.~\ref{system_arch}(c), except with the center hole diameter $d2=0.6$ mm, different from the $d2=1.0$ mm used in the main text. While a smaller hole delivers a larger $B_{1}$ field (hence increasing $\Omega_{\mathrm{R}}$), a shorter $T_{2}^{\mathrm{Rabi}}$ is expected due to the increased inhomogeneity of the $B_{1}$ field at the center of the hole. The higher $\Omega_{\mathrm{R}}$ of the HPHT diamond compared to that of the CVD diamond can be explained by the different projections of the $B_{1}$ vector onto the \nv axes being measured.\cite{B1Dir} }
        \label{comp_cvd_hpht}
\end{figure*}

\twocolumngrid

\begin{acknowledgments}
We would like to thank Jean-Philippe Tetienne for useful discussions, and Ye Kuang, Meilin Song, and Yiwen Zhang for their contributions. We acknowledge support from the School of Electrical Engineering and Telecommunications at UNSW Sydney, and the Australian Research Council (CE170100012). H.R.F. acknowledges the support of an Australian Government Research Training Program Scholarship. J.J.P. is supported by an Australian Research Council Discovery Early Career Research Award (DE190101397).
\end{acknowledgments}

\bibliography{mybib}

\begin{thebibliography}{10}
\bibitem{AntennaRes}
\bibinfo{note}{For ordering of the PCB antennas, it is important to choose the
  same substrate material (FR4) and substrate thickness (0.254~mm) as these
  parameters, and any other modifications in the antenna dimensions will result
  in a shift in the resonant frequency. A network analyzer can be used to
  determine the S$_{11}$ minimum of the antenna and hence its resonance
  frequency. Alternatively, the ODMR peak intensity or linewidth can be
  carefully monitored near $\sim$2.5~GHz to determine the exact resonant
  frequency.}
\bibitem{OscRot}
\bibinfo{note}{The factor of $\tfrac{1}{2}$ comes from the fact that we are
  using an oscillating, not a rotating, magnetic field.}
\end{thebibliography}

\end{document}